\documentclass[11pt]{article}
\usepackage[dvips]{graphicx}
\usepackage{amssymb}
\usepackage{amsmath}
\usepackage{amscd}
\usepackage{mathrsfs}
\usepackage{latexsym}
\usepackage{makeidx}

\numberwithin{equation}{section}
\date{}
\author{Sandro Sozzo\footnote{E-mail: sozzo@le.infn.it} \ \ and \ 
Claudio Garola\footnote{E-mail: garola@le.infn.it} \\ Dipartimento di Fisica and Sezione INFN \\ Universit\`a del Salento, via Arnesano, 73100 Lecce, Italy}
\title{\textbf{A Hilbert Space Representation of Generalized Observables and Measurement Processes in the ESR Model}}
\begin{document}
\maketitle

\begin{abstract}
\noindent
The \emph{extended semantic realism} (\emph{ESR}) \emph{model} recently worked out by one of the authors embodies the mathematical formalism of standard (Hilbert space) quantum mechanics in a noncontextual framework, reinterpreting quantum probabilities as \emph{conditional} instead of \emph{absolute}. We provide here a Hilbert space representation of the generalized observables introduced by the ESR model that satisfy a simple physical condition, propose a generalization of the projection postulate, and suggest a possible mathematical description of the measurement process in terms of evolution of the compound system made up of the measured system and the measuring apparatus.

\vspace{.3cm}

\vspace{.3cm}
\end{abstract}

\section{The ESR model \label{modello}}
The \emph{extended semantic realism} (\emph{ESR}) model has been proposed by one of the authors together with other authors to show that, contrarily to a widespread belief, the mathematical formalism of standard (Hilbert space) quantum mechanics (QM) can be embodied in a noncontextual framework \cite{ga03,gp04,g07}. We refer to \cite{gs08} for a detailed description of the ESR model and only recall here some of its features that are needed in the following.

According to the ESR model, every \emph{physical system} $\Omega$ is characterized by a set $\mathcal S$ of \emph{states}, a set $\mathcal O$ of \emph{generalized observables} and a set $\mathcal E$ of \emph{microscopic properties}. 

Each state $S \in {\mathcal S}$ is operationally defined as a class of physically equivalent \emph{preparing devices} \cite{bc81,l83}. Every preparing device $\pi$, when constructed and activated, performs a preparation of an individual example $x$ of $\Omega$ (\emph{physical object}), and one briefly says that ``$x$ is in the state $S$'' if $\pi$ belongs to $S$. 

Each generalized observable $A_0 \in {\mathcal O}$ is operationally defined as a class of physically equivalent \emph{measuring apparatuses}, and it is obtained in the ESR model by considering an observable $A$ of QM with set of possible values $\Xi$ on the real line $\Re$ and adding a further outcome $a_0 \in \Re \setminus \Xi$ (\emph{no--registration outcome} of $A_0$), so that the set of all possible values of $A_0$ is $\Xi_0=\Xi \cup \{ a_0 \}$.\footnote{One assumes here, for the sake of simplicity, that $\Re \setminus \Xi$ is non--void. This assumption is not restrictive. Indeed, if $\Xi=\Re$, one can choose a bijective Borel function $f: \Re \rightarrow \Xi'$ such that $\Xi' \subset \Re$ and replace $A$ by $f(A)$.}

Finally, microscopic properties play the role of theoretical entities (hence, they have no direct physical interpretation) and are such that, for every physical object $x$, every $f \in \mathcal E$ either is possessed or it is not possessed by $x$, independently of any measurement procedure. The set of microscopic properties possessed by a given physical object $x$ defines its \emph{microscopic state} $S^{i}$ which also plays the role of a theoretical entity.

Let now $\mathbb{B}(\Re)$ be the $\sigma$--algebra of all Borel subsets of $\Re$. The set ${\mathcal F}_{0}$ of all \emph{macroscopic properties} of $\Omega$ is defined by
\begin{equation}
{\mathcal F}_{0} \ =  \ \{ (A_0, X) \ | \  A_0 \in {\mathcal O}, \ X \in \mathbb{B}(\Re) \},
\end{equation}
and the subset ${\mathcal F} \subset {\mathcal F}_{0}$ is defined by
\begin{equation}
{\mathcal F} \ =  \ \{ (A_0, X) \ | \  A_0 \in {\mathcal O}, \ X \in \mathbb{B}(\Re), \ a_0 \notin X \}.
\end{equation}

Then, one assumes that a bijective mapping $\varphi: {\mathcal E} \rightarrow {\mathcal F}$ exists. By using this assumption one can provide a description of an \emph{idealized measurement} of a macroscopic property $F=(A_0, X)$ on a physical object $x$ in the state $S$, that is supposed to be performed by a dichotomic registering device which yields outcome \emph{yes} if the value of $A_0$ belongs to $X$, \emph{no} otherwise. Whenever $F \in \mathcal F$ one gets the fundamental equation of the ESR model
\begin{equation} \label{formuladipartenza}
p_{S}^{t}(F)=p_{S}^{d}(F)p_{S}(F),
\end{equation}
where $p_{S}^{t}(F)$ is the overall probability that the measurement yield the yes outcome when $F$ is measured on $x$, $p_{S}^{d}(F)$ is the probability that $x$ be detected, and $p_{S}(F)$ is the conditional probability that the measurement yield the yes outcome when $x$ is detected.

Eq. (\ref{formuladipartenza}) has been extensively discussed in \cite{gs08}. We recall here that the \emph{detection probability} $p_{S}^{d}(F)$ does not depend on features of the measuring apparatus nor is influenced by the environment because Eq. (\ref{formuladipartenza}) applies to idealized measurements only (which are the counterpart in the ESR model of the \emph{ideal first kind measurements} of standard QM). The ESR model assumes instead that $p_{S}^{d}(F)$ depends on the microscopic properties possessed by $x$, because these properties may be such that the no--registration outcome $a_0$ occurs even if an idealized measurement is performed. This assumption is introduced as a theoretical hypothesis that can be confirmed or falsified by testing its empirical consequences, and is not based on an underlying description (\emph{e.g.}, wave or particle) of physical objects.\footnote{It is interesting to note that a wave model has been recently provided according to which \emph{unfair sampling} occurs when considering a measurement process in which the measuring apparatus has a threshold \cite{a09}. We have proven in \cite{gs08} that an unconventional kind of unfair sampling occurs in the ESR model and explains the predicted violation of Bell's inequalities. Hence one may wonder whether also this unfair sampling can be justified by using the foregoing wave model to describe the physical objects that are considered. But the answer is negative, because no space of parameters (``hidden variables'') associated with the measuring apparatuses occurs in the ESR model whenever idealized measurements are considered. The reader can refer to \cite{gs08} for a more detailed treatment of this topic and a brief comparison of the ESR model with the V\"{a}xj\"{o} interpretation of QM \cite{k07}.}

Making reference to Eq. (\ref{formuladipartenza}), the basic assumption of the ESR model can be stated as follows.

\vspace{.2cm}
\noindent
\emph{Whenever S is a pure state, $p_{S}(F)$ can be evaluated by using the same rules that yield the probability of $F$ in the state $S$ according to QM.} 
\vspace{.2cm}

The above assumption implies that the ESR model incorporates the mathematical formalism of QM and its rules for calculating probabilities, but interprets such rules as providing \emph{conditional} (with respect to detection) instead of \emph{absolute} probabilities. As a consequence, the ESR model yields some predictions that are formally identical to those of QM but have a different physical interpretation, and further predictions that differ also formally from those of QM \cite{g07,gs08,s07}. The ESR model thus constitutes a new theoretical scheme. At this stage, however, the mathematical representation of the physical entities that are introduced in it is only partial, and a formal treatment of the detection probabilities is lacking, though some predictions on such probabilities can already be obtained. The next section will therefore be devoted to start a research on these topics.

\section{Hilbert space formalism for the ESR model \label{generalizedformalism}}
According to the ESR model the probability $p_{S}(F)$ in Eq. (\ref{formuladipartenza}) can be evaluated as in QM, associating in particular the physical system $\Omega$ with a (separable) complex Hilbert space $\mathscr H$ and representing every pure state $S$ of $\Omega$ by a unit vector of $\mathscr H$ or by a one--dimensional orthogonal projection operator. We adopt this representation as a general representation of physical systems and pure states in the ESR model from now on. But, then, a generalized observable $A_0$ cannot be represented by a self--adjoint operator on $\mathscr H$, hence neither by a projection operator valued, or \emph{PV}, measure. We intend to show in this section that a simple Hilbert space representation of $A_0$ can be given if $A_0$ belongs to a special class of generalized observables, and that this representation leads to a straightforward generalization of the projection postulate.

Let us firstly recall that the generalized observable $A_0$ is obtained by considering an observable $A$ of QM and adding a no--registration outcome $a_0$ to the set $\Xi$ of all possible outcomes of $A$ (Sect. \ref{modello}). Therefore, let us introduce the symbol $\widehat{A}$ to denote the self--adjoint operator representing $A$ in QM (the spectrum of which obviously coincides with $\Xi$) and the symbol $P^{\widehat{A}}$ to denote the PV measure associated with $\widehat{A}$ by the spectral theorem, that is,
\begin{equation}
P^{\widehat{A}}: X \in \mathbb{B}(\Re) \longmapsto P^{\widehat{A}}(X) \in {\mathscr L}({\mathscr H}),
\end{equation}
where ${\mathscr L}({\mathscr H})$ is the set of all orthogonal projection operators on ${\mathscr H}$, $\widehat{A}=\int_{-\infty}^{+\infty} \lambda \mathrm{d} P^{\widehat{A}}_{\lambda}$, $\int_{-\infty}^{+\infty}  \mathrm{d} P^{\widehat{A}}_{\lambda}=I$, and, for every $X \in \mathbb{B}(\Re)$, $P^{\widehat{A}}(X)= \int_{X} \mathrm{d} P^{\widehat{A}}_{\lambda}$.

Let us now come to $A_0$. For the sake of simplicity, we assume here that $A_0$ satisfies the following condition.

\vspace{.2cm}
\noindent
\emph{C}. \emph{The detection probability $p_{S}^{d}(F)$ of a macroscopic property $F=(A_0, X) \in {\mathcal F}$ depends on $A_0$ but not on $X$}.
\vspace{.2cm}

Because of condition C we can write $p_{S}^{d}(A_0)$ in place of $p_{S}^{d}(F)$. Hence we obtain from Eq. (\ref{formuladipartenza})
\begin{equation} \label{overall1}
p_{S}^{t}((A_0, X))=p_{S}^{d}(A_0)p_{S}((A_0, X)).
\end{equation}
The probability $p_{S}^{t}(A_0, \{ a_0 \})$ of getting the outcome $a_0$ is instead given by
\begin{equation} \label{overall2}
p_{S}^{t}((A_0, \{ a_0 \}))=1-p_{S}^{d}(A_0).
\end{equation}
Moreover, the overall probability that a measurement of a macroscopic property $F=(A_0, X) \in {\mathcal F}_{0}\setminus {\mathcal F}$ on a physical object $x$ in the state $S$ yield the yes outcome is
\begin{equation} \label{overall3}
p_{S}^{t}((A_0, X))=p_{S}^{t}((A_0, X \setminus \{ a_0 \}))+p_{S}^{t}((A_0, \{ a_0 \})).
\end{equation}
Let $S$ be a pure state represented by the unit vector $|\psi\rangle$. Then, we put $p_{\psi}^{d}(\widehat{A})\equiv p_{S}^{d}(A_0)$. Furthermore, if $F=(A_0, X) \in {\mathcal F}$, we get, because of the basic assumption of the ESR model,
\begin{equation}
p_{S}((A_0, X))=\langle\psi|P^{\widehat{A}}(X)|\psi\rangle.
\end{equation}
Hence Eq. (\ref{overall1}) becomes
\begin{equation} \label{overallmath1}
p_{S}^{t}((A_0, X))=\langle\psi|p_{\psi}^{d}(\widehat{A})P^{\widehat{A}}(X)|\psi\rangle
\end{equation}
and Eq. (\ref{overall2}) becomes
\begin{equation} \label{overallmath2}
p_{S}^{t}((A_0, \{ a_0 \}))=1-p_{\psi}^{d}(\widehat{A})=\langle\psi|(1-p_{\psi}^{d}(\widehat{A}))I|\psi\rangle.
\end{equation}
In addition, let $F=(A_0, X) \in {\mathcal F}_{0}\setminus {\mathcal F}$. Since $P^{\widehat{A}}(X \setminus \{ a_0 \})=P^{\widehat{A}}(X)$, Eq. (\ref{overall3}) becomes 
\begin{equation} \label{overallmath3}
p_{S}^{t}((A_0, X))=\langle\psi|((1-p_{\psi}^{d}(\widehat{A}))I+p_{\psi}^{d}(\widehat{A})P^{\widehat{A}}(X))|\psi\rangle.
\end{equation}
Eqs. (\ref{overallmath1})--(\ref{overallmath3}) suggest one to associate with $A_0$, for every unit vector $|\psi \rangle \in \mathscr H$, a mapping
\begin{equation}
T_{\psi}^{\widehat{A}}: X \in \mathbb{B}(\Re) \longmapsto T_{\psi}^{\widehat{A}}(X) \in {\mathscr B}({\mathscr H}),
\end{equation}
where ${\mathscr B}({\mathscr H})$ denotes the set of all bounded linear operators on $\mathscr H$, defined by setting
\begin{equation} \label{POV}
T_{\psi}^{\widehat{A}}(\{ a_0 \})=(1-p_{\psi}^{d}(\widehat{A}))I,
\end{equation}
and, for every $X \in \mathbb{B}(\Re)$,
\begin{equation}
T_{\psi}^{\widehat{A}}(X) = \left \{
\begin{array}{cll} 
p_{\psi}^{d}(\widehat{A}) P^{\widehat{A}}(X) & & \textrm{if} \ a_0 \notin X \\
T_{\psi}^{\widehat{A}}(\{ a_0 \})+T_{\psi}^{\widehat{A}}(X \setminus \{ a_0 \})=(1-p_{\psi}^{d}(\widehat{A}))I+p_{\psi}^{d}(\widehat{A})P^{\widehat{A}}(X) & & \textrm{if} \ a_0 \in X
\end{array}
\right. .
\end{equation}
It follows immediately that, for every unit vector $|\psi\rangle \in \mathscr H$,

\vspace{.2cm}
(i) $T_{\psi}^{\widehat{A}}(\Re)=I$;

(ii) for every $X \in \mathbb{B}(\Re)$, $0 \le T_{\psi}^{\widehat{A}}(X) \le I$;

(iii) $T_{\psi}^{\widehat{A}}(\bigcup_{i} X_i)=\sum_{i}T_{\psi}^{\widehat{A}}(X_i)$, for every disjoint sequence $\{ X_i \in \mathbb{B}(\Re)  \}_{i}$
(where the series converges in the weak topology of ${\mathscr B}({\mathscr H})$).

\vspace{.2cm}
Hence, for every unit vector $|\psi\rangle \in \mathscr H$, $T_{\psi}^{\widehat{A}}$ is a positive operator valued, or \emph{POV}, measure. Moreover the following \emph{commutativity property} is satisfied,
\begin{center}
for every $X, Y \in \mathbb{B}(\Re)$, $T_{\psi}^{\widehat{A}}(X)T_{\psi}^{\widehat{A}}(Y)=T_{\psi}^{\widehat{A}}(Y)T_{\psi}^{\widehat{A}}(X)$.
\end{center}

Because of the above properties the generalized observable $A_0$ can be represented by the \emph{family of} (\emph{commutative}) \emph{POV measures}
\begin{equation}
\left \{ T_{\psi}^{\widehat{A}}: X \in \mathbb{B}(\Re) \longmapsto T_{\psi}^{\widehat{A}}(X) \in {\mathscr B}({\mathscr H}) \right \}_{|\psi\rangle\in {\mathscr H}, \parallel|\psi\rangle\parallel=1}.
\end{equation}
Indeed, bearing in mind Eqs. (\ref{overallmath1}), (\ref{overallmath2}) and (\ref{overallmath3}), one immediately gets that the probability that the outcome of a measurement of $A_0$ on a physical object $x$ in the state $S$ represented by the unit vector $|\psi\rangle$ belong to the Borel set $X$ is given by
\begin{equation} \label{gen_prob_psi}
p_{S}^{t}((A_0, X))=\langle\psi|T_{\psi}^{\widehat{A}}(X)|\psi\rangle,
\end{equation}
or, equivalently,
\begin{equation} \label{gen_prob_W}
p_{S}^{t}((A_0, X))=Tr [W_{\psi}T_{\psi}^{\widehat{A}}(X)]
\end{equation}
where $W_{\psi}=|\psi\rangle\langle\psi|$.\footnote{If the representation of the generalized observables satisfying condition C introduced here is compared with the representation of observables introduced by unsharp QM \cite{d76,blm91,bgl96,bl96,bs96,b98} two basic differences leap out.

\hspace*{.2cm}(i) A generalized observable satisfying condition C is represented by a \emph{family} of POV measures parametrized by the set of all vectors representing pure states, while an observable of unsharp QM is represented by a \emph{single} POV measure.

\hspace*{.2cm}(ii) Only commutative POV measures appear in the representation of a generalized observable satisfying condition C.

Difference (i) is especially relevant since it implies that the generalized observables introduced by the ESR model do not coincide, in general, with the observables  introduced by unsharp QM. This can be intuitively explained by recalling that the occurrence of the no--registration outcome when measuring a generalized observable depends only on intrinsic features of the physical object that is considered (microscopic properties), hence it neither depends on the measuring apparatus nor it has an unsharp source. Difference (ii) is less relevant, because it depends on the fact that only idealized measurements are considered in the ESR model (which correspond to sharp measurements in unsharp QM) and it would disappear in an unsharp extension of the ESR model.}

The mathematical representation provided above naturally leads to inquire into the state transformation induced by a nondestructive idealized measurement of a macroscopic property associated with a generalized observable satisfying condition C. If one accepts that measurements of this kind are minimally perturbing, one can assume that, if the state $S$ of a physical object $x$ is pure, it is not altered whenever $x$ is not detected, while it is modified according to standard QM rules whenever $x$ is detected. These assumptions lead one to introduce the following \emph{generalized projection postulate}.

\vspace{.2cm}
\noindent
\emph{GPP}. \emph{Whenever a nondestructive idealized measurement of a physical property $F=(A_0,X)\in {\mathcal F}_{0}$ is performed on a physical object $x$ in a pure state $S$ represented by the unit vector $|\psi\rangle$ or, equivalently, by the one--dimensional projection operator $W_{\psi}=|\psi\rangle\langle\psi|$, and the yes outcome is obtained, the state of $x$ after the measurement is a pure state $S_F$ represented by the unit vector}
\begin{equation} \label{projpostulate_psi}
|\psi_{F} \rangle=\frac{T_{\psi}^{\widehat{A}}(X)|\psi\rangle}{\sqrt{\langle\psi | T_{\psi}^{\widehat{A} \dag}(X)T_{\psi}^{\widehat{A}}(X) |\psi\rangle}} \ , 
\end{equation}
\emph{or, equivalently, by the one--dimensional orthogonal projection operator}
\begin{equation} \label{projpostulate_W}
W_{\psi_{F}}=\frac{T_{\psi}^{\widehat{A}}(X)W_{\psi}T_{\psi}^{\widehat{A} \dag}(X)}{Tr[W_{\psi}T_{\psi}^{\widehat{A} \dag}(X)T_{\psi}^{\widehat{A}}(X)]}.
\end{equation}
\emph{Furthermore, if the no outcome is obtained, Eqs. (\ref{projpostulate_psi}) and (\ref{projpostulate_W}) still hold with $\Re\setminus X$ in place of $X$.}

\vspace{.2cm}

GPP replaces the projection postulate, as stated in elementary textbooks and manuals on QM, introducing two basic changes. Firstly, the operator $T_{\psi}^{\widehat{A}}(X)$ takes the place of the projection operator that appears in the projection postulate. Secondly, the term under square root in Eq. (\ref{projpostulate_psi}) and the term in the denominator in Eq. (\ref{projpostulate_W}) do not coincide with the probability provided by Eq. (\ref{gen_prob_psi}) or (\ref{gen_prob_W}).   

In order to illustrate the content of GPP let us consider some particular cases.

(i) $X=\{a_0 \}$. If the measurement yields the yes outcome, then $|\psi_{F}\rangle=|\psi\rangle$, consistently with our assumptions above.

(ii) $a_0 \notin X$. If the measurement yields the yes outcome, then $|\psi_{F}\rangle=\frac{P^{\widehat{A}}(X)|\psi\rangle}{\sqrt{\langle\psi|P^{\widehat{A}}(X)|\psi\rangle}}$, consistently with our assumptions above.

(iii) $a_0 \in X$. If the measurement yields the yes outcome, then $|\psi_{F}\rangle= \alpha  |\psi\rangle  +  \beta  P^{\widehat{A}}(X)  |\psi\rangle$, where  $\alpha$ and $\beta$ are coefficients whose calculation is straightforward.

\vspace{.2cm}

Summing up, we conclude that we have obtained in this section a Hilbert space representation of a subclass of generalized observables in the ESR model which allows one to predict probabilities of outcomes of measurements (Eqs. (\ref{gen_prob_psi}) and (\ref{gen_prob_W})) and states after measurements (Eqs. (\ref{projpostulate_psi}) and (\ref{projpostulate_W})) whenever pure states only are considered.\footnote{GPP refers to nondestructive idealized measurements, hence one may wonder whether such kind of measurements can be classified as \emph{ideal measurements of the first kind} according to standard definitions in QM. Let us therefore suppose that a first measurement of $A_0$ is performed on a physical object $x$ in the state $S$ and then repeated on $x$ in the final state. If the first measurement yields outcome $a_n \ne a_0$, the second could yield $a_n$ as well as $a_0$; if the first measurement yields outcome $a_0$, the second could yield $a_n \ne a_0$ if the detection probability of $A_0$ in the state $S$ is not 0. Strictly speaking, the measurement is not a first kind measurement. It can be observed, however, that if the first measurement yields outcome $a_n$, the second can never yield outcome $a_m$, with $0 \ne m \ne n$. In this sense, we say that our measurement is a \emph{generalized} measurement of the first kind. Moreover, the outcome of the measurement determines the final state of the physical object $x$. In this sense, we say that our measurement is a generalized ideal measurement. Summarizing, we say that our nondestructive idealized measurements are \emph{generalized ideal measurements of the first kind}.}  Of course, these results demand a generalization to arbitrary generalized observables and mixed states, which we will not undertake in this paper. We note, however, that our above formalism has been adopted in order to facilitate this generalization.

\section{Discrete generalized observables and the measurement process}
We intend to show in this section that Eqs. (\ref{gen_prob_psi})--(\ref{projpostulate_W}) can be rewritten in a form which is closer to the corresponding equations of QM whenever discrete generalized observables satisfying condition C are considered. We also want to show that GPP can be justified in this case by introducing a suitable evolution of the physical system made up of the (microscopic) physical object plus the (macroscopic) measuring apparatus.

Let therefore $A_0$ be a generalized observable satisfying condition C and obtained from a discrete observable $A$ of QM represented by a self--adjoint operator $\widehat{A}$ whose spectrum is $\Xi= \{ a_1, a_2, \ldots \}$, which implies that the set of possible outcomes of $A_0$ is $\Xi_0= \{a_0, a_1, a_2, \ldots \}$. Let $P_{1}^{\widehat{A}}$, $P_{2}^{\widehat{A}}$, \ldots be the (orthogonal) projection operators associated with $a_1$, $a_2$, \ldots, respectively, by the spectral decomposition of $\widehat{A}$, and let us introduce, for every unit vector $|\psi\rangle \in \mathscr H$, a set ${\mathcal M}_{\psi}^{A_0}=\{ M_{\psi k}^{\widehat{A}} \}_{k \in {\mathbb N}_{0}}$ of \emph{generalized measurement operators} defined as follows
\begin{eqnarray}
M_{\psi 0}^{\widehat{A}} & = & \sqrt{1-p_{\psi}^{d}(\widehat{A})} \  I \ , \\
M_{\psi k}^{\widehat{A}} & = & \sqrt{p_{\psi}^{d}(\widehat{A})} P_{k}^{\widehat{A}} \ \ (k \in {\mathbb N}) 
\end{eqnarray}
(note that all the operators in ${\mathcal M}_{\psi}^{A_0}$ are linear, bounded, self--adjoint and positive; moreover, ${\mathcal M}_{\psi}^{A_0}$ is complete because $\sum_{k \in \mathbb N_0} M_{\psi k}^{\widehat{A} \dag}M_{\psi k}^{\widehat{A}}=I$, and commutative, because, for every $k,l \in {\mathbb N}_{0}$, $[M_{\psi k}^{\widehat{A}}, M_{\psi l}^{\widehat{A}}]=0$). By using the operators in ${\mathcal M}_{\psi}^{A_0}$ the overall probability that a physical object $x$ in the state $S$ represented by the unit vector $|\psi\rangle$ yield outcome $a_k$ ($k \in {\mathbb N}_{0}$) whenever $A_0$ is measured on it is given by
\begin{equation} \label{prob_dis_psi}
p_{S}^{t}((A_0, \{a_k \}))=\langle \psi | M_{\psi k}^{\widehat{A}\dag}M_{\psi k}^{\widehat{A}}| \psi\rangle 
\end{equation}
because of Eq. (\ref{gen_prob_psi}), or, equivalently, by
\begin{equation} \label{prob_dis_W}
p_{S}^{t}((A_0, \{a_k \}))=Tr[W_\psi M_{\psi k}^{\widehat{A} \dag}M_{\psi k}^{\widehat{A}}] 
\end{equation}
because of Eq. (\ref{gen_prob_W}). Moreover, if the measurement yields outcome $a_k$, the state of $x$ after the measurement is represented by the unit vector
\begin{equation} \label{genpost_dis_psi}
|\psi_{k} \rangle=\frac{M_{\psi k}^{\widehat{A}}|\psi\rangle}{\sqrt{\langle\psi | M_{\psi k}^{\widehat{A} \dag}M_{\psi k}^{\widehat{A}}|\psi\rangle}} 
\end{equation}
because of Eq. (\ref{projpostulate_psi}) or, equivalently, by the one--dimensional orthogonal projection operator
\begin{equation} \label{genpost_dis_W}
W_{k}=\frac{M_{\psi k}^{\widehat{A}}W_{\psi}M_{\psi k}^{\widehat{A} \dag}}{Tr[W_{\psi}M_{\psi k}^{\widehat{A} \dag}M_{\psi k}^{\widehat{A}}]}
\end{equation}
because of Eq. (\ref{projpostulate_W}). The term under square root in Eq. (\ref{genpost_dis_psi}) and the term in the denominator in Eq. (\ref{genpost_dis_W}) now coincide with the probability provided by Eq. (\ref{prob_dis_psi}) or (\ref{prob_dis_W}).

 Eq. (\ref{genpost_dis_W}) implies, in particular, that if the measurement is \emph{nonselective} the final state of $x$ is a mixed state represented in standard QM by the density operator
\begin{equation}\label{nonsel_gpp_k}
W=\sum_{k \in {\mathbb N}_{0}} Tr[W_{\psi}M_{\psi k}^{\widehat{A} \dag} M_{\psi k}^{\widehat{A}}] W_k=\sum_{k \in {\mathbb N}_{0}} M_{\psi k}^{\widehat{A}}W_{\psi}M_{\psi k}^{\widehat{A} \dag}.
\end{equation}

Let us suppose now, for the sake of simplicity, that the spectrum $\Xi$ of $\widehat{A}$ is not only discrete but also nondegenerate, and put $|\psi\rangle=\sum_{k \in {\mathbb N}} c_k | a_k \rangle$ (where $| a_k \rangle$ is the eigenvector associated with the eigenvalue $a_k$ of $\widehat{A}$). Then Eq. (\ref{nonsel_gpp_k}) yields
\begin{eqnarray} \label{entstate_W}
W=[1-p_{\psi}^{d}(\widehat{A})]|\psi\rangle\langle\psi|+p_{\psi}^{d}(\widehat{A})\sum_{k \in {\mathbb N}} P_{k}^{\widehat{A}}|\psi\rangle\langle\psi|P_{k}^{\widehat{A}}= \nonumber \\
=[1-p_{\psi}^{d}(\widehat{A})]|\psi\rangle\langle\psi|+p_{\psi}^{d}(\widehat{A})\sum_{k \in {\mathbb N}} |c_k|^{2} | a_k \rangle\langle a_k |. \label{genprojpostulate}
\end{eqnarray}

Eq. (\ref{genprojpostulate}) can be used to justify GPP in the special case of discrete generalized observables satisfying condition C by assuming a suitable evolution of the compound system made up of the microscopic measured object and the macroscopic measuring apparatus. Indeed, let us consider the apparatus measuring $A_0$ as an individual example of a macroscopic physical system $\Omega_M$ associated with the Hilbert space ${\mathscr H}_{M}$. Let $|1 \rangle, \ldots, |k \rangle, \ldots$ be the unit vectors of ${\mathscr H}_{M}$ representing the macroscopic states of $\Omega_M$ which correspond to the outcomes $a_1, \ldots, a_k, \ldots$, respectively. Moreover, let us introduce the unit vector $| 0 \rangle$ which represents the macroscopic state of the apparatus when it is ready to perform a measurement or when the physical object $x$ is not detected. Finally, let us assume that $\{| 0 \rangle, |1 \rangle, \ldots, |k \rangle, \ldots \}$ is an orthonormal basis in ${\mathscr H}_{M}$. Let $S_0$ be the initial state of the compound system made up of the physical object $x$ plus the macroscopic apparatus, represented by the unit vector $|\psi\rangle |0\rangle$, and let us assume the following (generally nonlinear, hence nonunitary\footnote{A unitary evolution has been recently proposed by one of us together with other authors \cite{gp04}. Our present assumption seems to fit better with the general view of the ESR model.}) time evolution of the compound system
\begin{equation}
\begin{CD}
|\psi\rangle |0\rangle =\sum_{k \in {\mathbb N}} c_k | a_k \rangle |0\rangle @>>  > \alpha_{\psi} \sum_{k \in {\mathbb N}} c_k | a_k \rangle | k \rangle + \beta_{\psi} |\psi\rangle |0\rangle,
\end{CD}
\end{equation}
where, for every unit vector $|\psi\rangle$, $\alpha_{\psi}, \beta_{\psi} \in \mathbb C$ and $\alpha_{\psi}=\sqrt{p_{\psi}^{d}(\widehat{A})}e^{i \theta_{\psi}}$, $\beta_{\psi}=\sqrt{1-p_{\psi}^{d}(\widehat{A})}e^{i \varphi_{\psi}}$, hence $|\alpha_{\psi}|^{2}+|\beta_{\psi}|^{2}=1$.

Let us now consider the density operator $W_{C}$ associated with the final state of the compound system after the interaction. $W_C$ can be written as 
\begin{displaymath}
W_{C}= (\alpha_{\psi} \sum_{k \in {\mathbb N}} c_k | a_k \rangle | k\rangle + \beta_{\psi} |\psi\rangle |0\rangle) (\alpha_{\psi}^{*} \sum_{l \in {\mathbb N}} c_l^{*} \langle a_l | \langle l| + \beta^{*}_{\psi} \langle\psi| \langle 0|)=
\end{displaymath}
\begin{displaymath}
=|\alpha_{\psi}|^{2}\sum_{k,l \in {\mathbb N}} c_k c_l^{*} | a_k \rangle \langle a_l | \otimes | k \rangle \langle l |+\alpha_{\psi} \beta^{*}_{\psi} \sum_{k \in {\mathbb N}} c_k |a_k\rangle\langle\psi| \otimes | k\rangle\langle 0 |+
\end{displaymath}
\begin{equation}
+ \alpha_{\psi}^{*} \beta_{\psi} \sum_{l \in {\mathbb N}} c_l^{*}|\psi\rangle\langle a_l | \otimes | 0 \rangle\langle l|+|\beta_{\psi}|^{2}|\psi\rangle\langle\psi| \otimes | 0 \rangle \langle 0|.
\end{equation}
The final state of the measured physical object $x$ can be represented in standard QM by the density operator obtained by performing the partial trace of $W_{C}$ with respect to ${\mathscr H}_{M}$,
\begin{equation} \label{partrac}
Tr_{M} W_{C}=[1-p_{\psi}^{d}(\widehat{A})]|\psi\rangle\langle\psi|+p_{\psi}^{d}(\widehat{A})\sum_{k \in {\mathbb N}} |c_k|^{2} | a_k \rangle\langle a_k|.
\end{equation}
The second term of Eq. (\ref{partrac}) coincides with the second term of Eq. (\ref{genprojpostulate}), which provides the desired justification of GPP.

It is important to observe that the justification above is complete, in the sense that also the interpretations of the two descriptions provided by Eqs. (\ref{genprojpostulate}) and (\ref{partrac}) coincide. Indeed, because of objectivity of physical properties, all probabilities in Eq. (\ref{partrac}) are epistemic according to the ESR model, exactly as the probabilities in Eq. (\ref{entstate_W}), which does not occur in QM or its unsharp extension \cite{bs96,b98,gs07a}.


\begin{thebibliography}{99}

\bibitem{ga03} Garola, C.: Embedding quantum mechanics into an objective framework. Found. Phys. Lett. \textbf{16}, 605--612 (2003)

\bibitem{gp04} Garola, C., Pykacz, J.: Locality and measurements within the SR model for an objective interpretation of quantum mechanics. Found. Phys. \textbf{34}, 449--475 (2004)

\bibitem{g07} Garola, C.: The ESR model: reinterpreting quantum probabilities in a realistic and local framework. In: Adenier, G., et al. (eds) Quantum Theory: Reconsideration of Foundations 4, pp. 247--252. American Institute of Physics, Melville, New York (2007)

\bibitem{gs08} Garola, C., Sozzo, S.: Embedding quantum mechanics into a broader noncontextual theory: a conciliatory result. ArXiv:0811.0539v2 [quant-ph] (2009). Submitted to Int. J. Theor. Phys.

\bibitem{bc81} Beltrametti, E.G., Cassinelli, G.: The Logic of Quantum Mechanics. Addison--Wesley, Reading, MA (1981)

\bibitem{l83} Ludwig, G.: Foundations of Quantum Mechanics I. Springer, Berlin (1983)

\bibitem{a09} Adenier, G.: Violation of Bell inequalities as a violation of fair sampling in threshold detectors. In: Accardi L., \emph{et al.} (eds) Foundations of Probability and Physics-5, pp. 8--17. American Institute of Physics, Melville, New York (2009)

\bibitem{k07} Khrennikov A.Y.: Contextual Approach to Quantum Formalism. Springer, Berlin (2009)

\bibitem{s07} Sozzo, S.: Modified BCHSH inequalities within the ESR model. In: Adenier, G., et al. (eds) Quantum Theory: Reconsideration of Foundations 4, pp. 334--338. American Institute of Physics, Melville, New York (2007)

\bibitem{d76} Davies, E.B.: Quantum Theory of Open Systems. Academic Press, London (1976)



\bibitem{blm91} Busch, P., Lahti, P.J., Mittelstaedt, P.: The Quantum Theory of Measurement. Springer, Berlin (1991)

\bibitem{bgl96} Busch, P., Grabowski, M., Lahti, P.J.: Operational Quantum Physics. Springer, Berlin (1996)

\bibitem{bl96} Busch P., Lahti, P.J.: The standard model of quantum measurement theory. Found. Phys. \textbf{26}, 875--893 (1996)

\bibitem{bs96} Busch P., Shimony, A.: Insolubility of the quantum measurement problem for unsharp observables. Stud. His. Phil. Mod. Phys. \textbf{27B}, 397--404 (1996) 

\bibitem{b98} Busch, P.: Can `unsharp objectification' solve the quantum measurement problem?. Int. J. Theor. Phys. \textbf{37}, 241--247 (1998)

\bibitem{gs07a} Garola, C., Sozzo, S.: The physical interpretation of partial traces: two nonstandard views. Theor. Math. Phys. \textbf{152}(2), 1087--1098 (2007)

\end{thebibliography}
\end{document}